\documentclass[reprint,aps,pra,superscriptaddress,showpacs,amsmath,amssymb,floatfix]{revtex4-1}

\usepackage{txfonts}

\allowdisplaybreaks[4]

\usepackage{graphicx}

\usepackage{subfigure}
\usepackage{braket}
\usepackage{bm}

\def\articletitle#1{{\it #1.}}

\usepackage{color}

\graphicspath{{./}}

\newcommand{\nature}{ Nature }
\newcommand{\natphoton}{ Nat.\ Photon.\ }
\newcommand{\science}{ Science }
\renewcommand{\prl}{ Phys.\ Rev.\ Lett.\ }

\renewcommand{\pra}{ Phys.\ Rev.\ A }

\newcommand{\optexp}{ Opt.\ Exp.\ }

\newcommand{\abs}[1]{\ensuremath{\left\vert#1\right\vert}}

\newcommand{\sciap}{ Light\ Science\ and\ Applications\ }

\newcommand{\UTokyo}{
Department of Applied Physics, School of Engineering, \\
The University of Tokyo, 7-3-1 Hongo, Bunkyo-ku, Tokyo 113-8656, Japan}
\newcommand{\UTokyoQP}{
JST, PRESTO, 4-1-8 Honcho, Kawaguchi, Saitama, 332-0012, Japan}

\begin{document}

\title{Complete temporal mode characterization of non-Gaussian states\\ by dual homodyne measurement}

\author{Kan Takase}
\email{takase@alice.t.u-tokyo.ac.jp}
\affiliation{\UTokyo}
\author{Masanori Okada}
\affiliation{\UTokyo}
\author{Takahiro Serikawa}
\affiliation{\UTokyo}
\author{Shuntaro Takeda}
\affiliation{\UTokyo}
\affiliation{\UTokyoQP}
\author{Jun-ichi Yoshikawa}
\affiliation{\UTokyo}
\author{Akira Furusawa}
\email{akiraf@ap.t.u-tokyo.ac.jp}
\affiliation{\UTokyo}

\date{\today}

\begin{abstract}
Optical quantum states defined in temporal modes, especially non-Gaussian states like photon-number states, play an important role in quantum computing schemes. In general, the temporal-mode structures of these states are characterized by one or more complex functions called temporal-mode functions (TMFs). Although we can calculate TMF theoretically in some cases, experimental estimation of TMF is more advantageous to utilize the states with high purity. In this paper, we propose a method to estimate complex TMFs. This method can be applied not only to arbitrary single-temporal-mode non-Gaussian states but also to two-temporal-mode states containing two photons. This method is implemented by continuous-wave (CW) dual homodyne measurement and doesn't need prior information of the target states nor state reconstruction procedure. We demonstrate this method by analyzing several experimentally created non-Gaussian states.
\end{abstract}

\pacs{03.67.-a,42.50.Dv,42.50.Ex}

\maketitle

\section{Introduction}
Quantum states of light are a promising resource of quantum computation \cite{Knill.nature(2001),Menicucci.prl(2006),Takeda.prl(2017)} and quantum communication \cite{Abruzzo.pra(2013),Brask.prl(2010)}. They are characterized by optical modes like polarization, spatial, and temporal modes. Among these modes, temporal modes have a lot of flexibility thus are useful for many applications. One prominent example is temporal-mode multiplexing of quantum states to realize large-scale fault-tolerant quantum computation \cite{Takeda.prl(2017),Menicucci.pra(2011),Menicucci.prl(2014),Takeda.nature(2013),Yoshikawa.qph,Yokoyama(2013),Yoshikawa.apl}. In this scheme, grasping temporal-mode structures of quantum states are essential for basic operations like interference and measurement. Therefore, a methodology to characterize the states' temporal-mode is in great demand.

A temporal mode $f$ is characterized by a complex function $f(t)$, called temporal-mode function (TMF). For example, single-photon states in a temporal mode $f$ are given by $\ket{1_f}\equiv \hat{a}_f^{\dag} \ket{\tilde{0}}$, where $\ket{\tilde{0}}$ is a multi-mode vacuum state and $\hat{a}_f^{\dag} \equiv \int dt f(t)\hat{a}^{\dag}(t)$ is a creation operator of temporal-mode $f$. Such non-Gaussian states, which can be used as ancillary states or quantum information carrier in quantum computation, are the main interest of temporal-mode analysis. Although we can calculate TMFs theoretically in some state creation schemes \cite{Anne.pra(2007)}, imperfection of experiment varies their actual forms. This mismatch leads to extra photon loss in useful applications. Therefore, experimental estimation of TMFs of optical quantum states is essential to utilize the states with high purity.

So far, Single-temporal-mode states, the states described by one temporal mode, have been actively analyzed experimentally. Especially, continuous wave (CW) homodyne and heterodyne measurements are powerful tools. In Refs. \cite{MacRae.prl(2012),Morin.prl(2013),Yoshikawa.prx(2013)}, the TMFs of single-photon states and Schr\"{o}dinger's cat states are estimated by applying principal components analysis (PCA) \cite{Adbi(2010)}, which decomposes correlated variables into uncorrelated variables, to CW homodyne signals. This method doesn't require prior information of the target states and can access the TMFs without a state reconstruction algorithm. We can apply this method to arbitrary single-temporal-mode non-Gaussian states. However, their TMFs are limited to real functions, because PCA gives us only real functions. Ref. \cite{Qin.sciap(2015)} estimates complex TMFs of single-photon states by constructing what they call temporal density matrix via CW heterodyne tomography. In the case of general non-Gaussian states, however, we need to consider larger size density matrix having higher-photon number components, and estimation of TMFs is not trivial. On top of that, we need several local oscillator (LO) beams having different frequencies for construction of temporal density matrix.

\begin{figure}[t]
\centering
\includegraphics[width=\linewidth]{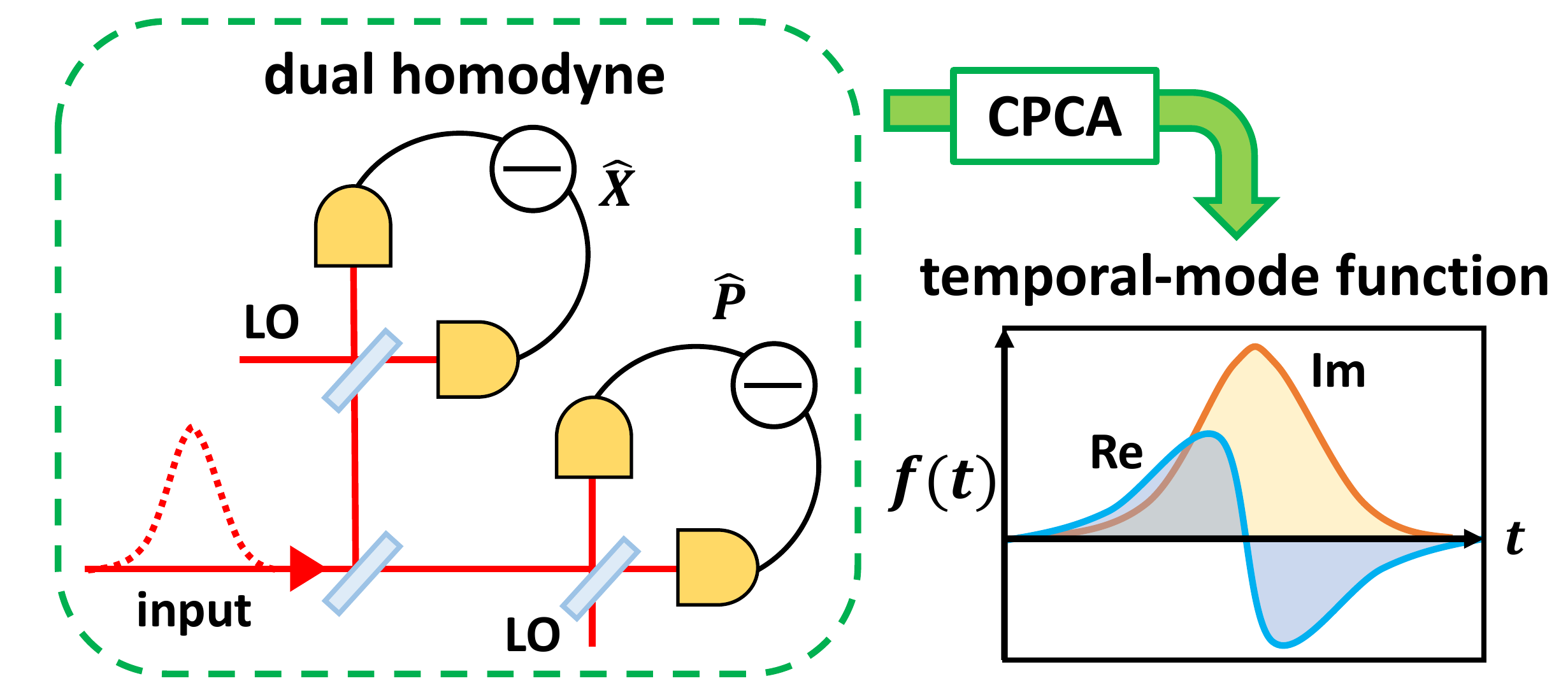}
\caption{
Conceptual diagram. The input state is a non-Gaussian state in unknown temporal modes. We estimate the temporal mode structure by measuring conjugate quadratures $\hat{X}, \hat{P}$ by dual homodyne measurement, and processing the data by CPCA.
}
\label{CPCA}
\end{figure}

In this paper, we propose a method what we call {\it complex-number PCA (CPCA)}; estimation of complex TMFs by applying PCA to complex variables given by CW dual homodyne measurement. Figure \ref{CPCA} is a conceptual diagram of our method. CPCA can deal with arbitrary single-temporal-mode non-Gaussian states characterized by complex TMFs.  On top of that, it can be applied to dual-temporal mode states containing two photons $\hat{a}_{f_1}^{\dag}\hat{a}_{f_2}^{\dag}\ket{\tilde{0}}$ to estimate the complex TMFs $f_1(t),f_2(t)$. These states are the simplest example of multi-temporal-mode states. Although multi-temporal mode states play an important role in applications \cite{Chuang.pra(1996),Wasilewaki.pra(2007),Bergmann.pra(2016)}, this kind of analysis has not been done in previous researches. Our method possibly opens a way to TMF estimation of general multi-temporal mode states. Like previous PCA method, CPCA needs neither prior information, state reconstruction procedure nor LO beams having different frequencies. The simplicity and capability of our method to characterize wide range of quantum states would lead to useful applications not only in state creation experiments but also in quantum communication and quantum computation schemes \cite{Knill.nature(2001),Menicucci.prl(2006),Takeda.prl(2017),Abruzzo.pra(2013),Brask.prl(2010)}. We experimentally demonstrate this method using several non-Gaussian states characterized by complex TMFs.

This paper is organized as follows. In Sec. \ref{2a}, we define temporal modes of light and optical single- and multi-temporal-mode quantum states. In Sec. \ref{2b}, we review previous estimation of real TMFs by PCA. In Sec. \ref{2c}, we discuss applying PCA to CW dual homodyne measurement signals. In Sec. \ref{2d}, we discuss how to analyze two-photon states $\hat{a}_{f_1}^{\dag}\hat{a}_{f_2}^{\dag}\ket{\tilde{0}}$. In Sec. \ref{2e}, we discuss how photon loss affects our analysis. In Sec. \ref{3a}, we review a creation method of non-Gaussian states. In Sec. \ref{3b}, we explain the experimental setup. In Sec. \ref{3c}, we show the experimentally estimated TMFs of several non-Gaussian states.


\section{THEORY}\label{22}
\subsection{Definition of temporal modes\label{2a}}
We introduce optical TMFs and basic operators. A temporal mode $f$ is characterized by time spectrum $f(t)$ called TMF, in which quantum states can be defined. The TMFs are complex functions in general. We define the instantaneous annihilation and creation operators $\hat{a}(t)$ and $\hat{a}^{\dag}(t)$ which satisfy the commutation relation $[\hat{a}(t),\hat{a}^{\dag}(t')]=\delta(t-t')$ where $\delta(t)$ is the Dirac delta function. Then, photon annihilation and creation in a temporal mode $f$ are described by operators given by
\begin{eqnarray}
\hat{a}_f \equiv \int dt f^*(t) \hat{a}(t)\ ,\ 
\hat{a}_f^{\dag} \equiv \int dt f(t) \hat{a}^{\dag}(t).
\label{eq:modeopdef}
\end{eqnarray}
Suppose complex function $f(t)$ satisfies $\int dt \abs{f(t)}^2=1$ so that these operators satisfy $[\hat{a}_f,\hat{a}_f^{\dag}]=1$. In general, 
\begin{eqnarray}
[\hat{a}_{f_j},\hat{a}_{f_k}^{\dag}] &=& \int dt \int dt'\ f_j^*(t)f_k(t') [\hat{a}(t),\hat{a}^{\dag}(t')] \nonumber \\
&=& \int dt \ f_j^*(t)f_k(t)  \nonumber \\
&\equiv& \braket{f_j , f_k}.
\label{eq:commutation}
\end{eqnarray}
Therefore, the inner product of TMFs gives the commutation relation of temporal modes.

In experiments, we often treat physical quantities in finite and discrete time. For that, we define time-bin modes $\{ t_j \}_{j=1}^M$ which divide time interval $[0,T]$ into small $M$ time-bins. Their TMFs are given by
\begin{eqnarray}
t_j(t) = \begin{cases}
    \sqrt{M/T} & ((j-1)\cdot T/M  \le t < j \cdot T/M) \\
    0  & ({\rm otherwise}) \ .
\label{eq:timebinmode}
  \end{cases}
\end{eqnarray}
In many cases, it is useful to use annihilation and creation operators of time-bin modes
\begin{eqnarray}
\hat{a}_{t_j} = \int dt \ t_j(t) \hat{a}(t)\ ,\ 
\hat{a}_{t_j}^{\dag} = \int dt \ t_j(t) \hat{a}^{\dag}(t) \ ,
\end{eqnarray}
instead of instantaneous operators $\hat{a}(t),\hat{a}^{\dag}(t)$. Their commutation relation is given by $[\hat{a}_{t_j},\hat{a}_{t_k}^{\dag}] = \delta_{j,k}$.

When $f(t)$ has finite values only in $[0,T]$ and varies slowly during the time change $T/M$, it is well approximated as
\begin{eqnarray}
f(t) \approx \sum_{j=1}^M f[t_j]t_j(t) ,\label{eq:fdiscreteapprox}
\end{eqnarray}
where $f[t_j] \equiv \sqrt{T/M}f(j\cdot T/M)$. Strictly speaking, the both sides of Eq. (\ref{eq:fdiscreteapprox}) are not the same, but in the rest of this paper, we use an equal sign in the following for convenience,
\begin{eqnarray}
f(t) = \sum_{j=1}^M f[t_j]t_j(t) . \label{eq:fdiscrete}
\end{eqnarray}
Then, by using time-bin modes, annihilation and creation operators of temporal mode $f$ are given by
\begin{eqnarray}
\hat{a}_f = \sum_{j=1}^M f^*[t_j] \ \hat{a}_{t_j}\ ,\ 
\hat{a}_f^{\dag} = \sum_{j=1}^M f[t_j] \ \hat{a}_{t_j}^{\dag} . \label{eq:afdis}
\end{eqnarray}
Their commutation relation is given by
\begin{eqnarray}
[\hat{a}_{f_j},\hat{a}_{f_k}^{\dag}] = \sum_{l=1}^M f_j^*(t_l)f_k(t_l) .
\end{eqnarray}
In this way, we can describe temporal mode $f$ both in infinite-continuous time and finite-discrete time.

When quantum states are described only by creation operators of temporal mode $f$ and background vacuum state $\ket{\tilde{0}}$, we call them single-temporal-mode states in temporal mode $f$. Photon-number states $\ket{n_f}\equiv \frac{1}{\sqrt{n!}}\left(\hat{a}_f^{\dag}\right)^n\ket{\tilde{0}}$ are the examples of such states, and they are orthogonal complete basis of single-temporal-mode states in temporal mode $f$. On the other hand, we need more than one temporal mode to describe multi-temporal-mode states. The basic examples are multi-temporal-mode Fock states $\hat{a}_{f_1}^{\dag}\hat{a}_{f_2}^{\dag}\cdots\hat{a}_{f_n}^{\dag}\ket{\tilde{0}}$. We underline that TMFs $\{ f_j(t) \}_{j=1}^n$ are arbitrary complex functions and not orthogonal to each other in general, thus we cannot describe the states as $\ket{1_{f_1},1_{f_2},\cdots,1_{f_n}}$ except for the special case when $\{ f_j(t) \}_{j=1}^n$ are orthogonal functions. Our goal is to estimate TMFs of those states by experimentally deciding the discrete values $\{ f[t_j] \}_{j=1}^M$ in Eq. (\ref{eq:fdiscrete}).

\subsection{Estimation of real TMFs\label{2b}}
Our goal is to estimate complex TMFs $f(t)$. In this section, however, we concentrate on the case where $f(t)$ are real functions. Such functions can be estimated by applying principal components analysis (PCA) to quadratures as introduced in Refs. \cite{MacRae.prl(2012),Morin.prl(2013),Yoshikawa.prx(2013)}. We extend this method to complex number in Sec. \ref{2c}.
\subsubsection{Data acquisition}
When we try to estimate $f(t)$, one useful information is quadrature values. The quadratures of instantaneous modes are defined by
\begin{eqnarray}
\hat{x}_{\theta}(t) \equiv \frac{\hat{a}(t)\mathrm{e}^{-i\theta}+\hat{a}^{\dag}(t)\mathrm{e}^{i\theta}}{\sqrt{2}} \ .
\end{eqnarray}
Ideal homodyne detector can measure the quadrature $\hat{x}_{\theta}(t)$ using CW coherent beam called LO having a phase $\theta$. The quadratures of a temporal mode $f$ is given by
\begin{eqnarray}
\hat{x}_{f,\theta} \equiv \frac{\hat{a}_f\mathrm{e}^{-i\theta}+\hat{a}_f^{\dag}\mathrm{e}^{i\theta}}{\sqrt{2}} = \int dt \  f(t) \hat{x}_{\theta}(t) \ . \label{eq:xf}
\end{eqnarray}
Note that $f(t)$ is a real function here. We can obtain $\hat{x}_{f,\theta}$ by integrating the results of ideal CW homodyne measurement with a weight $f(t)$. In the rest of the paper, we omit the suffix $\theta$ except when it is important.

In PCA, the mode conversion of quadratures given by Eq. (\ref{eq:xf}) plays a central role, thus we need precise value of $\hat{x}(t)$. Unfortunately, homodyne detectors in laboratories are not ideal, preventing us from knowing the exact values of $\hat{x}(t)$. When the impulse response of the measurement system is given by $g(t)$, CW homodyne measurement signals are given by (under proper normalization)
\begin{eqnarray}
\hat{x}'(t) = \int dt' \ \hat{x}(t')g(t-t') \ .
\end{eqnarray}
In the following, we assume $g(t)$ is mainly determined by low-pass filter effect due to the finite bandwidth of the homodyne detector. Then, quadratures of temporal-mode $f$ we can obtain are given by
\begin{eqnarray}
\hat{x}'_{f} \equiv \int dt \  f(t) \hat{x}'(t) = \int dt \  \hat{x}(t) \int dt' f(t')g(t'-t) \ .
\end{eqnarray}
When $g(t)$ is narrow enough compared to $f(t)$, that is, the homodyne detector is broadband enough compared to the bandwidth of $F(\omega)\equiv \int dt f(t)\mathrm{e}^{-i\omega t}$, we can regard $g(t)$ as a delta function $\delta(t)$. It follows that
\begin{eqnarray}
\hat{x}'_{f} = \int dt \  \hat{x}(t) f(t) = \hat{x}_{f} \ .
\end{eqnarray}
Therefore, even when our homodyne detectors are not ideal, we can neglect the imperfection if they are broadband enough compared to $F(\omega)$. In the following, we express $\hat{x}'(t)$ and $\hat{x}'_{f}$ just as $\hat{x}(t)$ and $\hat{x}_{f}$ for convenience, thus $\hat{x}(t)$ and $\hat{x}_{f}$ means measured and calculated values by using homodyne detectors having finite but broad enough bandwidth.

In experiments, we measure quadrature values at appropriate sampling rate during reasonable time span. Thus, we have to treat finite and discrete time. Let us assume $f(t)$ has finite values only in time $[0,T]$. Then, we measure quadrature values $M$ times at the same interval during $[0,T]$. Here, the sampling rate $M/T$ should be large enough that $f(t)$ and $\hat{x}(t)$ vary slowly in time change $T/M$. Note that homodyne measurement is an observation in a system rotating at the carrier frequency of LO beam, thus the time change of $f(t)$ and $\hat{x}(t)$ are not so fast. In this situation, time-bin modes $\{ t_j \}_{j=1}^M$ introduced in Eq. (\ref{eq:timebinmode}) are useful. From above and Eq. (\ref{eq:xf}), the quadratures of time-bin modes are given by
\begin{eqnarray}
\hat{x}_{t_j} = \int dt \  t_j(t) \hat{x}(t) \approx \sqrt{T/M} \ \hat{x}(j\cdot T/M) .
\end{eqnarray}
Similarly to Eq. (\ref{eq:fdiscrete}), we use an equal sign in the following,
\begin{eqnarray}
\hat{x}_{t_j} = \sqrt{T/M} \ \hat{x}(j\cdot T/M) .
\end{eqnarray}
Thus, we can determine $\hat{x}_{t_j}$ by finite sampling rate homodyne measurement. From Eq. (\ref{eq:afdis}), quadratures of temporal-mode $f$ is given by
\begin{eqnarray}
\hat{x}_{f} = \sum_{j=1}^{M} f[t_j]\ \hat{x}_{t_j} \ . \label{eq:discrete}
\end{eqnarray}
This equation corresponds to Eq. (\ref{eq:xf}). Like above, we can define and measure quadratures both in infinite-continuous time and finite-discrete time.

\subsubsection{Principal component analysis}
PCA \cite{Adbi(2010)} is an analysis procedure to convert correlated variables into a set of uncorrelated variables called principal components. In PCA, the first principal component has the largest variance in the whole space, and following components are decided to have  the largest variance in the subspace which is orthogonal to the components decided before.

In quantum optics, PCA has been applied to the quadratures $\{\hat{x}_{t_j}\}_{j=1}^M$ to estimate real TMFs $f(t)$ of non-Gaussian states \cite{MacRae.prl(2012),Morin.prl(2013),Yoshikawa.prx(2013)}. The variables $\{\hat{x}_{t_j}\}_{j=1}^M$ are correlated because autocorrelation functions $\braket{\hat{x}_{t_j}\hat{x}_{t_k}}$ are not zero in general when $j \neq k$. PCA is a procedure to find uncorrelated variables $\{\hat{x}_{e_j}\}_{j=1}^M$ satisfying
\begin{eqnarray}
\braket{\hat{x}_{e_j}\hat{x}_{e_k}} &=& \braket{\hat{x}_{e_j}^2} \delta_{j,k} \ , \\
\braket{\hat{x}_{e_1}^2} \ge \braket{\hat{x}_{e_2}^2} &\ge& \cdots \ge \braket{\hat{x}_{e_M}^2} . \label{eq:uncor}
\end{eqnarray}
The functions $\{ e_j(t) \}_{j=1}^M$ have important information about the TMFs we want to estimate.

For example, when we apply PCA to single-photon states $\hat{a}_f^{\dag}\ket{\tilde{0}}$,
we get $\braket{\hat{x}_{e_1}^2}=3/2 \ , \ \braket{\hat{x}_{e_2}^2}=\cdots =\braket{\hat{x}_{e_M}^2}=1/2$, and $f(t) = e_1(t)$ \cite{Morin.prl(2013)}. This is because single-photon states have three times larger quadrature variance than that of vacuum states in uncorrelated modes. Generally, there exists a certain phase where single-temporal-mode states, typically non-Gaussian states, have larger variance of quadratures than other uncorrelated vacuum modes. Therefore, $e_1(t)$ is expected to be the TMF of the single-temporal mode states if we choose a proper phase of LO.

We can carry out PCA by introducing a matrix $V_t$ given by
\begin{eqnarray}
V_t = \left(
    \begin{array}{cccc}
      \braket{\hat{x}_{t_1}^2} & \braket{\hat{x}_{t_1}\hat{x}_{t_2}} & \ldots & \braket{\hat{x}_{t_1}\hat{x}_{t_M}} \\
      \braket{\hat{x}_{t_2}\hat{x}_{t_1}} & \braket{\hat{x}_{t_2}^2} & \ldots & \braket{\hat{x}_{t_2}\hat{x}_{t_M}} \\
      \vdots & \vdots & \ddots & \vdots \\
      \braket{\hat{x}_{t_M}\hat{x}_{t_1}} &\braket{\hat{x}_{t_M}\hat{x}_{t_2}} & \ldots &\braket{\hat{x}_{t_M}^2}
    \end{array}
  \right) .
\end{eqnarray}
We can obtain this matrix via CW homodyne measurement. When we measure a target state with a sampling rare $M/T$ during $[0,T]$, we get a set of quadratures $\{ \hat{x}_{t_j} \}_{j=1}^M$. By repeating the same measurement, we can calculate the average values $\braket{\hat{x}_j\hat{x}_k}$. $V_t$ is a real symmetrical matrix thus diagonalized as $EV_tE^T$ by a certain orthogonal matrix $E$. In this case, such matrix $E$ is given by
\begin{eqnarray}
E = \left(
    \begin{array}{cccc}
      e_1[t_1] & e_1[t_2] & \ldots & e_1[t_M] \\
      e_2[t_1] & e_2[t_2] & \ldots & e_2[t_M] \\
      \vdots & \vdots & \ddots & \vdots \\
      e_M[t_1] & e_M[t_2] & \ldots & e_M[t_M]
    \end{array}
  \right) .
\end{eqnarray}
Why $EV_tE^T$ is diagonalized is shown as follows. $EV_tE^T$ is given by
\begin{eqnarray}
EV_tE^T = 
\left(
    \begin{array}{cccc}
      \braket{\hat{x}_{e_1}^2} & \braket{\hat{x}_{e_1}\hat{x}_{e_2}} & \ldots & \braket{\hat{x}_{e_1}\hat{x}_{e_M}} \\
      \braket{\hat{x}_{e_2}\hat{x}_{e_1}} & \braket{\hat{x}_{e_2}^2} & \ldots & \braket{\hat{x}_{e_2}\hat{x}_{e_M}} \\
      \vdots & \vdots & \ddots & \vdots \\
      \braket{\hat{x}_{e_M}\hat{x}_{e_1}} &\braket{\hat{x}_{e_M}\hat{x}_{e_2}} & \ldots &\braket{\hat{x}_{e_M}^2}
    \end{array}
  \right) , \label{eq:ve}
\end{eqnarray}
where we use a relation that follows from Eq. (\ref{eq:discrete}),
\begin{eqnarray}
\braket{\hat{x}_{e_j}\hat{x}_{e_k}} = \sum_{l,m=1}^M \ e_j[t_l]\ e_k[t_m] \braket{\hat{x}_{t_l}\hat{x}_{t_m}} .
\end{eqnarray}
From Eq. (\ref{eq:uncor}), the off-diagonal terms of Eq. (\ref{eq:ve}) are zero,
\begin{eqnarray}
EV_tE^T &=& 
\left(
    \begin{array}{cccc}
      \braket{\hat{x}_{e_1}^2} & & & \\
      & \braket{\hat{x}_{e_2}^2} &  & \mbox{\strut\rlap{\smash{\huge$0$}}\quad} \\
      & & \ddots & \\
      \mbox{\strut\rlap{\smash{\huge$0$}}\quad}& &  &\braket{\hat{x}_{e_M}^2}
    \end{array}
  \right) \nonumber \\
&\equiv& {\rm diag}\left[ \braket{\hat{x}_{e_1}^2},\braket{\hat{x}_{e_2}^2}, \cdots , \braket{\hat{x}_{e_M}^2} \right] .
\end{eqnarray}
In this way, we can obtain eigenfunctions $\{ e_j(t) \}_{j=1}^M$ by getting matrix $V_t$ via CW homodyne measurement and calculating a matrix $E$ which diagonalizes $V_t$. Note that due to the orthogonality of $E$, the eigenfunctions $\{ e_j(t) \}_{j=1}^M$ are orthogonal,
\begin{eqnarray}
\braket{e_j,e_k} = \sum_{l=1}^M e_j[t_l]\ e_k[t_l] = \delta_{j,k} \ , \ 1 \le j,k \le M \ .
\end{eqnarray}
Especially, in single-temporal-mode state analysis, TMF $f(t)$ is given by
\begin{eqnarray}
f(t) = e_1(t) = \sum_{j=1}^M e_1[t_j]\ t_j(t) = \sum_{j=1}^M E_{1,j} \ t_j(t) \ .
\end{eqnarray}

PCA has been applied to experimentally created non-Gaussian states such as single-photon states and Schr\"{o}dinger's cat states in a single temporal mode \cite{MacRae.prl(2012),Morin.prl(2013),Yoshikawa.prx(2013)}. In this method, however, the eigenfunctions $\{ e_j(t) \}_{j=1}^M$ are limited to real functions because they are given by linear combination of real functions $\{ t_j(t) \}_{j=1}^M$ by orthogonal matrix $E$. Therefore, this method is not suitable for the quantum states which have complex TMFs. In the next section, we extend PCA to estimation of complex TMFs.

\subsection{Estimation of complex TMFs\label{2c}}
\subsubsection{Dual-homodyne measurement}
In the previous section, eigenfunctions $\{ e_j(t) \}_{j=1}^M$ are limited to real functions because we apply PCA to quadratures $\{ \hat{x}_{t_j} \}_{j=1}^M$, which are real numbers. In order to estimate complex TMFs, we have to apply PCA to complex variables. CW dual homodyne measurement and CW heterodyne measurement give such variables, measuring the conjugate quadratures $\hat{x}_{\theta}(t)$ and $\hat{x}_{\theta+\frac{\pi}{2}}(t)$ simultaneously. Actually, in Ref. \cite{Qin.sciap(2015)}, complex TMFs of single-photon states are estimated by CW heterodyne measurement using several LOs having different frequencies. In Ref. \cite{Qin.sciap(2015)}, however, an optimization algorithm is adopted, requiring heavier calculation than PCA.  In this section, we discuss applying PCA to CW dual homodyne measurement signals. Dual homodyne can be implemented by LOs having one frequency, unlike heterodyne method used in Ref. \cite{Qin.sciap(2015)}.

\begin{figure}[t]
\centering
\includegraphics[width=0.9\linewidth]{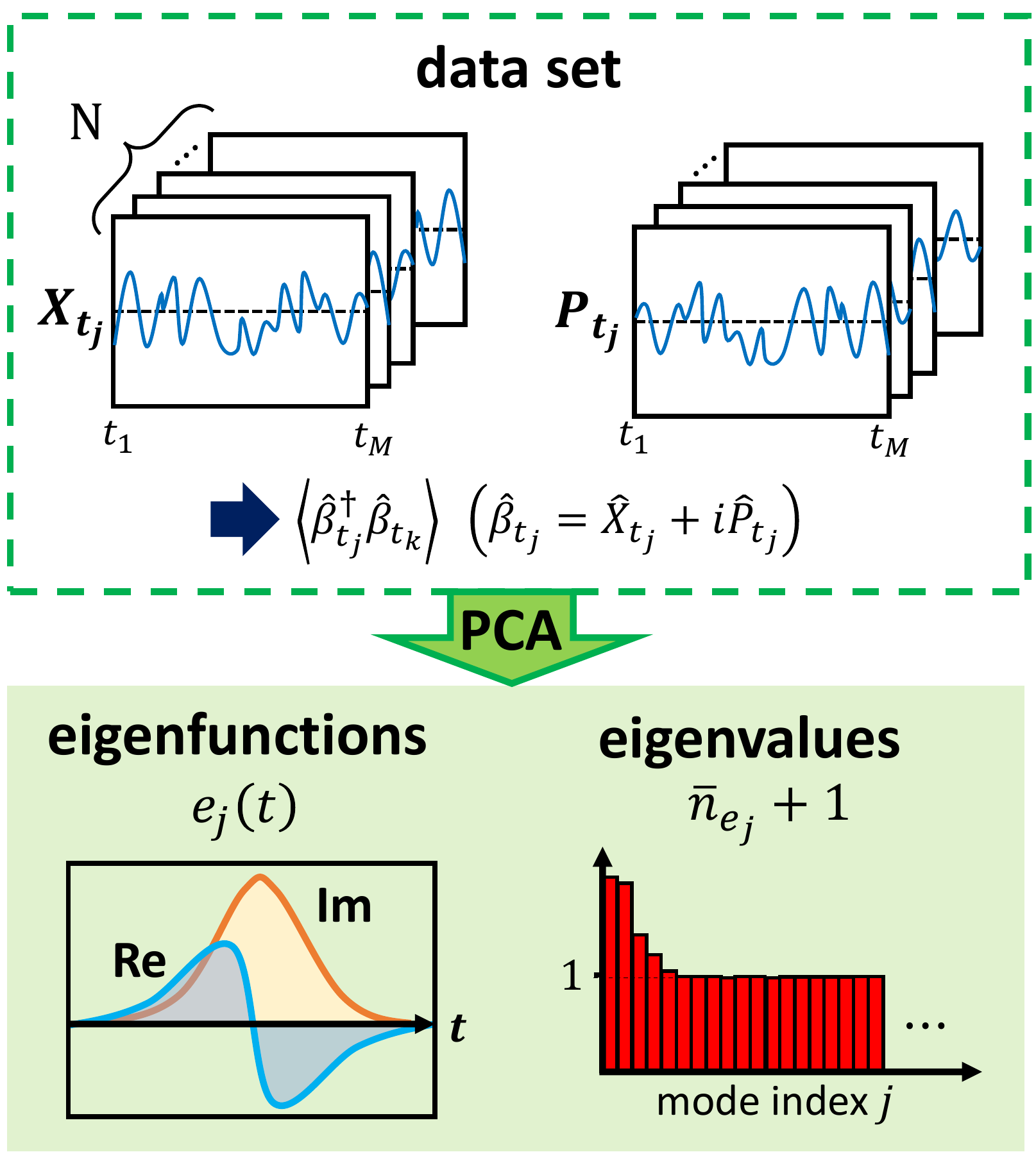}
\caption{
Schematic of CPCA. For every single shot measurement, we take the quadrature values $M$ times in $[0,T]$. We convert the complex variable $\hat{\beta}_{t_j}$ into uncorrelated variables by PCA, where we utilize the correlation $\braket{\hat{\beta}_{t_j}^{\dag}\hat{\beta}_{t_k}}$ calculated from $N$-frame data. From the eigenfunctions and eigenvalues, we can estimate the TMFs of the input states.
}
\label{fig:r2}
\end{figure}

To begin with, we explain CW dual homodyne measurement. Figure \ref{fig:r2} is a conceptual diagram of CW dual homodyne measurement. In this measurement, the target state $\ket{\Psi}$ (quadratures $\hat{x}_{\theta}(t)$) is divided into two outputs by a 50:50 beam splitter. Another input of the beam splitter is a vacuum state $\ket{0}$ (quadratures $\hat{x}_{\theta}^{(v)}(t)$). The two outputs are measured by CW homodyne detectors by using orthogonal phases of LOs. Measurement operators of these homodynes are given by
\begin{eqnarray}
\hat{X}_{t_j} = \frac{\hat{x}_{t_j}-\hat{x}_{t_j}^{(v)}}{\sqrt{2}} \ , \ \hat{P}_{t_j} = \frac{\hat{p}_{t_j}+\hat{p}_{t_j}^{(v)}}{\sqrt{2}} \ ,
\end{eqnarray}
where we put  $\hat{x} \equiv \hat{x}_{\theta = 0} ,\hat{p} \equiv \hat{x}_{\theta = \pi/2}$. We treat these values as a complex number $\hat{\beta}_{t_j}$ given by
\begin{eqnarray}
\hat{\beta}_{t_j} \equiv \hat{X}_{t_j}+i\hat{P}_{t_j} = \hat{a}_{t_j} - \hat{a}_{t_j}^{{\dag}(v)} \ .
\end{eqnarray}
This value corresponds to complex amplitude of the state $\ket{\Psi}$. The vacuum term $\hat{a}_{t_j}^{{\dag}(v)}$ means the uncertainty of simultaneous measurement of $\hat{x}_{t_j}$ and $\hat{p}_{t_j}$. We define similar value about temporal-mode $f$ as below,
\begin{eqnarray}
\hat{\beta}_{f} &\equiv& \sum_{j=1}^{M} f^*[t_j]\ \hat{\beta}_{t_j} = \hat{a}_f - \hat{a}_{f^*}^{\dag (v)} \ , \label{eq:beta} \\
\hat{\beta}_{f}^{\dag} &\equiv& \sum_{j=1}^{M} f[t_j]\ \hat{\beta}_{t_j}^{\dag} = \hat{a}_f^{\dag} - \hat{a}_{f^*}^{(v)} \ . \label{eq:betadag}
\end{eqnarray}
Note that $f(t)$ is a complex function, no longer a real function like in Sec. \ref{2b}. The distribution of $\hat{\beta}_{f_j}$ is given by the $Q$-function of the state in the mode $f$ \cite{Freyberger.pra(1993),Leonhardt.pra(1993)}.

\subsubsection{Complex-number principal components analysis}
Next, let us discuss applying PCA to the complex variables $\{ \hat{\beta}_{t_j} \}_{j=1}^M$. This process converts correlated variables $\{\hat{\beta}_{t_j}\}_{j=1}^M$ into uncorrelated variables $\{\hat{\beta}_{e_j}\}_{j=1}^M$ which satisfy
\begin{eqnarray}
\braket{\hat{\beta}_{e_j}^{\dag}\hat{\beta}_{e_k}} = \braket{\hat{\beta}_{e_j}^{\dag}\hat{\beta}_{e_j}} &\delta_{j,k}& \ , \\
\braket{\hat{\beta}_{e_1}^{\dag}\hat{\beta}_{e_1}} \ge \braket{\hat{\beta}_{e_2}^{\dag}\hat{\beta}_{e_2}} \ge \cdots &\ge& \braket{\hat{\beta}_{e_M}^{\dag}\hat{\beta}_{e_M}}.
\end{eqnarray}

Let us confirm the physical meaning of the value $\braket{\hat{\beta}_{f}^{\dag}\hat{\beta}_{f}}$. Equations (\ref{eq:beta}) and (\ref{eq:betadag}) lead to
\begin{eqnarray}
\hat{\beta}_{f}^{\dag}\hat{\beta}_{f} = \hat{a}_{f}^{\dag}\hat{a}_{f} +\hat{a}_{f^*}^{\dag(v)}\hat{a}_{f^*}^{(v)} +1 - \hat{a}_{f}^{\dag}\hat{a}_{f^*}^{\dag(v)} - \hat{a}_{f}\hat{a}_{f^*}^{(v)} .
\end{eqnarray}
Because CW dual-homodyne measurement has two-mode input $\ket{\Psi} \ket{0_v}$, the average value is given by
\begin{eqnarray}
\braket{\hat{\beta}_{f}^{\dag}\hat{\beta}_{f}} 
&=& \bra{0_v} \bra{\Psi}\hat{\beta}_{f}^{\dag}\hat{\beta}_{f} \ket{\Psi} \ket{0_v} \nonumber \\
&=& \bra{\Psi} \hat{a}_{f}^{\dag}\hat{a}_{f} \ket{\Psi} +1 \nonumber \\
&\equiv& \bar{n}_f+1 \ .  \label{eq:betatoa}
\end{eqnarray}
Therefore, $\braket{\hat{\beta}_{f}^{\dag}\hat{\beta}_{f}}$ shows the average photon number the state $\ket{\Psi}$ has in a temporal-mode $f$. For example, when applying PCA to single-photon states $\hat{a}_f^{\dag}\ket{\tilde{0}}$, we get $\bar{n}_{e_1}=1, \ \bar{n}_{e_2}=\cdots =\bar{n}_{e_M}=0$ and $e_1(t)=f(t)$. Generally, when we analyze single-temporal-mode states, only $\bar{n}_{e_1}$ is larger than zero, and $e_1(t)=f(t)$. To distinguish from previous PCA, we call this method as complex-number PCA (CPCA) in this paper.

We can carry out CPCA by introducing a matrix $C_t$ given by
\begin{eqnarray}
C_t = \left(
    \begin{array}{cccc}
      \braket{\hat{\beta}_{t_1}^{\dag}\hat{\beta}_{t_1}} & \braket{\hat{\beta}_{t_1}^{\dag}\hat{\beta}_{t_2}} & \ldots & \braket{\hat{\beta}_{t_1}^{\dag}\hat{\beta}_{t_M}} \\
      \braket{\hat{\beta}_{t_2}^{\dag}\hat{\beta}_{t_1}} & \braket{\hat{\beta}_{t_2}^{\dag}\hat{\beta}_{t_2}} & \ldots & \braket{\hat{\beta}_{t_2}^{\dag}\hat{\beta}_{t_M}} \\
      \vdots & \vdots & \ddots & \vdots \\
      \braket{\hat{\beta}_{t_M}^{\dag}\hat{\beta}_{t_1}} &\braket{\hat{\beta}_{t_M}^{\dag}\hat{\beta}_{t_2}} & \ldots &\braket{\hat{\beta}_{t_M}^{\dag}\hat{\beta}_{t_M}}
    \end{array}
  \right) . \label{eq:ct}
\end{eqnarray}
We can obtain this matrix via CW dual homodyne measurement. We measure a target state with a sampling rate $M/T$ during $[0,T]$ to get a set of values $\{ \hat{\beta}_{t_j} \}_{j=1}^M$. By repeating the same measurement, we can calculate $\braket{\hat{\beta}_{t_j}^{\dag}\hat{\beta}_{t_k}}$. $C_t$ is an Hermite matrix thus diagonalized by a unitary matrix $E$ as follows,
\begin{eqnarray}
E = \left(
    \begin{array}{cccc}
      e_1[t_1] & e_1[t_2] & \ldots & e_1[t_M] \\
      e_2[t_1] & e_2[t_2] & \ldots & e_2[t_M] \\
      \vdots & \vdots & \ddots & \vdots \\
      e_M[t_1] & e_M[t_2] & \ldots & e_M[t_M]
    \end{array}
  \right) , \label{eq:Ematrix}
\end{eqnarray}
\begin{eqnarray}
EC_tE^{\dag}
&=& {\rm diag}\left[\braket{\hat{\beta}_{e_1}^{\dag}\hat{\beta}_{e_1}} ,\braket{\hat{\beta}_{e_2}^{\dag}\hat{\beta}_{e_2}}, \cdots , \braket{\hat{\beta}_{e_M}^{\dag}\hat{\beta}_{e_M}} \right] \nonumber \\
&=& {\rm diag}\left[\bar{n}_{e_1}+1 ,\bar{n}_{e_2}+1, \cdots , \bar{n}_{e_M}+1 \right] \ . \label{eq:diagonalizationE}
\end{eqnarray}
In the diagonalization of $C_t$, we use a relation derived from Eqs. (\ref{eq:beta}) and (\ref{eq:betadag}),
\begin{eqnarray}
\braket{\hat{\beta}_{e_j}^{\dag}\hat{\beta}_{e_k}} = \sum_{l,m=1}^M \ e_j[t_l]\ e_k^*[t_m] \braket{\hat{\beta}_{t_l}^{\dag}\hat{\beta}_{t_m}} .
\end{eqnarray}
Like PCA, we can obtain eigenfunctions $\{ e_j(t) \}_{j=1}^M$ and average photon numbers $\{ \bar{n}_{e_j} \}_{j=1}^M$ through diagonalizing process. The important thing is that $\{ e_j(t) \}_{j=1}^M$ are complex functions because the matrix $E$ is unitary. It also follows
\begin{eqnarray}
\braket{e_j,e_k} = \delta_{j,k} \ , \ 1 \le j,k \le M . \label{eq:ortho}
\end{eqnarray}
In single-temporal-mode state analysis, TMF $f(t)$ is given by
\begin{eqnarray}
f(t) = e_1(t) = \sum_{j=1}^M e_1[t_j]\ t_j(t) = \sum_{j=1}^M E_{1,j} \ t_j(t) \ .
\end{eqnarray}
Like above, we can estimate complex TMF of single-temporal-mode states via CPCA.

\subsection{Dual-temporal-mode state analysis\label{2d}}
In the previous section, we discussed single-temporal-mode state analysis by CPCA. However, useful states are often defined in multi-temporal modes. For example, some quantum error correction codes use multi-mode states to protect fragile quantum information \cite{Chuang.pra(1996),Wasilewaki.pra(2007),Bergmann.pra(2016)}. Therefore, we should develop mode characterization tools for multi-mode states. As a simple case, we treat two photons distributed in two temporal modes. As shown in \cite{Yoshikawa.qph}, such states always can be given by
\begin{eqnarray}
\ket{\Psi_2} = \frac{1}{\sqrt{1+\abs{\braket{f_1,f_2}}^2}}\hat{a}_{f_1}^{\dag}\hat{a}_{f_2}^{\dag}\ket{\tilde{0}},
\label{eq:trit}
\end{eqnarray}
where $f_1(t), f_2(t)$ are complex functions not orthogonal in general. We will characterize the temporal-mode structures of $\ket{\Psi_2}$ by estimating  $f_1(t)$ and $f_2(t)$. In the following, we assume $f_1(t) \neq f_2(t)$.

When we apply CPCA to $\ket{\Psi_2}$, $\bar{n}_{e_2}$ in no longer zero, but has  positive value. Thus, $e_1$ and $e_2$ have the information of temporal-mode structures of $\ket{\Psi_2}$. From Eq. (\ref{eq:ortho}), the modes $e_1$ and $e_2$ are orthogonal. Note that in most cases $e_1, e_2$ are not the same modes as $f_1, f_2$, because $f_1, f_2$ are not orthogonal in general. The two photons in $\ket{\Psi_2}$ are distributed in $e_1,e_2$, thus $\ket{\Psi_2}$ can be described by
\begin{eqnarray}
\ket{\Psi_2} &=& \alpha\ket{2_{e_1},0_{e_2}} + \beta \ket{1_{e_1},1_{e_2}} + \gamma \ket{0_{e_1},2_{e_2}} \nonumber \\
&=& \left( \frac{\alpha}{\sqrt{2}} \hat{a}_{e_1}^{\dag \ 2} + \beta \ \hat{a}_{e_1}^{\dag}\hat{a}_{e_2}^{\dag} + \frac{\gamma}{\sqrt{2}} \hat{a}_{e_2}^{\dag \ 2} \right) \ket{\tilde{0}} , \label{eq:twophotonstates}
\end{eqnarray}
where $\alpha \in \mathbb{R}$ and $\beta,\gamma \in \mathbb{C}$ satisfy $\alpha^2+\abs{\beta}^2+\abs{\gamma}^2=1$. The quadratic polynomial of the creation operators in Eq. (\ref{eq:twophotonstates}) can be decomposed into a product of linear polynomials,
\begin{eqnarray}
\ket{\Psi_2} = \frac{\left( d_{11} \hat{a}_{e_1}^{\dag} + d_{12} \hat{a}_{e_2}^{\dag} \right)\left( d_{21} \hat{a}_{e_1}^{\dag} + d_{22} \hat{a}_{e_2}^{\dag} \right)}{\sqrt{1+\abs{d_{11}^*d_{21}+d_{12}^*d_{22}}^2}} \ket{\tilde{0}},
\end{eqnarray}
where $\abs{d_{11}}^2+\abs{d_{12}}^2=1,\ \abs{d_{21}}^2+\abs{d_{22}}^2=1$. From Eq. (\ref{eq:modeopdef}), we can use the next relation for arbitrary normalized orthogonal functions $g_1(t)$ and $g_2(t)$,
\begin{eqnarray}
\eta_1\hat{a}_{g_1}^{\dag} + \eta_2 \hat{a}_{g_2}^{\dag} = \hat{a}_{\eta_1g_1+\eta_2g_2}^{\dag}
\ \ \ \ \ (\abs{\eta_1}^2+\abs{\eta_2}^2=1) . \label{eq:conb}
\end{eqnarray}
Thus, we can get
\begin{eqnarray}
\ket{\Psi_2} = \frac{1}{\sqrt{1+\abs{d_{11}^*d_{21}+d_{12}^*d_{22}}^2}} \hat{a}_{d_{11}e_1+d_{12}e_2}^{\dag} \hat{a}_{d_{21}e_1+d_{22}e_2}^{\dag} \ket{\tilde{0}}. \nonumber \\
\end{eqnarray}
This is the same form as Eq. (\ref{eq:trit}), thus $f_1(t),f_2(t)$ are given by 
\begin{eqnarray}
\left(
    \begin{array}{c}
      f_1(t) \\
      f_2(t)
    \end{array}
  \right) 
=
\left(
    \begin{array}{cc}
      d_{11} & d_{12} \\
      d_{21} & d_{22}
    \end{array}
  \right) 
\left(
    \begin{array}{c}
      e_1(t) \\
      e_2(t)
    \end{array}
  \right)
\equiv D \left(
    \begin{array}{c}
      e_1(t) \\
      e_2(t)
    \end{array}
  \right) . \label{eq:eftrans}
\end{eqnarray}
Our goal is to calculate $f_1(t)$ and $f_2(t)$ experimentally. We can get $e_1(t), e_2(t)$ by CPCA, thus we express the matrix $D$ by experimentally obtainable values.

By assuming $d_{11},d_{21} \in \mathbb{R}$, we can determine $D$ uniquely from $\alpha,\beta,$ and $\gamma$. From Eqs. (\ref{eq:betatoa}) and (\ref{eq:twophotonstates}), these values should satisfy
\begin{eqnarray}
\bar{n}_{e_1} &=& \braket{\Psi_2| \hat{\beta}_{e_1}^{\dag}\hat{\beta}_{e_1} |\Psi_2} -1 = 2\alpha^2+|\beta|^2 , \nonumber \\
\bar{n}_{e_2} &=& \braket{\Psi_2| \hat{\beta}_{e_2}^{\dag}\hat{\beta}_{e_2} |\Psi_2} -1 = |\beta|^2+2|\gamma|^2 , \\
0 &=& \braket{\Psi_2| \hat{\beta}_{e_1}^{\dag}\hat{\beta}_{e_2} |\Psi_2} = \sqrt{2} \left( \alpha\beta + \beta^*\gamma \right). \nonumber  \label{eq:conditions}
\end{eqnarray}
Thus, when $\bar{n}_1>\bar{n}_2>0$,
\begin{eqnarray}
\alpha&=&\sqrt{\frac{\bar{n}_1}{2}} \ , \ \beta=0 \ , \ \abs{\gamma}=\sqrt{\frac{\bar{n}_2}{2}} . \label{answer1}
\end{eqnarray}
When $\bar{n}_1=\bar{n}_2=1$,
\begin{eqnarray}
\alpha&=&\abs{\gamma}=\sqrt{\frac{1-\abs{\beta}^2}{2}} \ , \ 2\arg{\beta}=\arg{\gamma} \pm \pi . \label{answer2}
\end{eqnarray}
There still exists uncertainty among $\alpha, \beta,$ and $\gamma$. Therefore, we cannot decide $D$ only from the CPCA results.

Interestingly, we can overcome this problem by introducing 4-th order moments. Firstly, we utilize
\begin{eqnarray}
q\mathrm{e}^{i\theta} \equiv \braket{\Psi_2 | \hat{\beta}_{e_1}^{\dag \ 2} \hat{\beta}_{e_2}^{2} | \Psi_2} = 2\alpha\gamma \ \ \ (q \ge 0) . \label{eq:qei}
\end{eqnarray}
You can easily obtain this value experimentally because you have the data set $\{ \hat{\beta}_{t_j} \}_{j=1}^M$ and eigenfunctions $e_1(t),e_2(t)$ to calculate the values $\hat{\beta}_{e_1}^{\dag}$, $\hat{\beta}_{e_2}$ using Eqs. (\ref{eq:beta}) and (\ref{eq:betadag}). Equation (\ref{eq:qei}) leads to
\begin{eqnarray}
2\alpha \cdot \abs{\gamma} = q \ , \ \arg{\gamma} = \theta . \label{addcond}
\end{eqnarray}
From Eqs. (\ref{answer1}), (\ref{answer2}) and (\ref{addcond}), we can determine $\alpha, \beta,$ and $\gamma$, thus the matrix $D$. When $\bar{n}_1>\bar{n}_2>0$,
\begin{eqnarray}
\alpha = \sqrt{\frac{\bar{n}_1}{2}} \ , \ \beta = 0 \ , \ \gamma = \sqrt{\frac{\bar{n}_2}{2}} \mathrm{e}^{i\theta} 
\end{eqnarray}
\begin{eqnarray}
D =
\frac{1}{\sqrt{\bar{n}_{e_1}^{\frac{1}{2}}+\bar{n}_{e_2}^{\frac{1}{2}}}}\left(
    \begin{array}{cc}
      \bar{n}_{e_1}^{\frac{1}{4}} \  & i \ \bar{n}_{e_2}^{\frac{1}{4}}\ \mathrm{e}^{i\frac{\theta}{2}} \\
      \bar{n}_{e_1}^{\frac{1}{4}} \  & -i \ \bar{n}_{e_2}^{\frac{1}{4}} \mathrm{e}^{i\frac{\theta}{2}}
    \end{array}
  \right) . \label{eq:linsp2}
\end{eqnarray}
When $\bar{n}_1=\bar{n}_2=1$,
\begin{eqnarray}
\alpha = \sqrt{\frac{q}{2}} \ , \ \beta = \pm i\sqrt{1-q}\mathrm{e}^{i\frac{\theta}{2}} \ , \ \gamma =  \sqrt{\frac{q}{2}} \mathrm{e}^{i\theta}
\end{eqnarray}

\begin{eqnarray}
D =
\left(
    \begin{array}{cc}
      \sqrt{1+\sqrt{1-q}} \  \ & \mp i \sqrt{1-\sqrt{1-q}}\ \mathrm{e}^{i\frac{\theta}{2}} \\
      \sqrt{1-\sqrt{1-q}} \  \ & \pm i \sqrt{1+\sqrt{1-q}}\ \mathrm{e}^{i\frac{\theta}{2}}
    \end{array}
  \right) . \label{Adeg}
\end{eqnarray}
The latter case corresponds to the situation when $f_1$ and $f_2$ are orthogonal, because the columns of $D$ are orthogonal. In this case, the phase of $\beta$ is still not unique. We can determine the phase by using another $4$-th order moment $\braket{\Psi_2 | \hat{\beta}_{e_1}^{\dag \ 2} \hat{\beta}_{e_1} \hat{\beta}_{e_2} | \Psi_2} = 2\alpha\beta$, which reveals $\arg{\beta}$.

In this way, we can characterize the temporal-mode structures of $\ket{\Psi_2}$ experimentally. What we have to do is executing CPCA to obtain $e_1(t), e_2(t), \bar{n}_{e_1},$ and $\bar{n}_{e_2}$, and calculating the $4$-th order moments from the dual homodyne signals and $e_1(t), e_2(t)$. Then, following Eqs. (\ref{eq:eftrans}) and (\ref{eq:linsp2}) or (\ref{Adeg}), we can calculate $f_1(t)$ and $f_2(t)$ in Eq. (\ref{eq:trit}). In the next section, we will discuss the case when the analysis objects have errors due to a lossy optical channel.

\subsection{Mixed states analysis\label{2e}}
We discussed pure states so far. In experiment, however, what we can prepare is mixed states due to the coupling between quantum states and the environment. Usually, the most dominant error is photon loss. In this section, we show that the analysis method discussed in Sec. \ref{2c} and \ref{2d} can work even when photon loss exists.

When we treat pure states $\ket{\Psi}$, the $(j,k)$ component of $C_t$ is given by
\begin{eqnarray}
C_{t,jk}\left(\ket{\Psi}\bra{\Psi}\right) = \bra{\Psi}\hat{\beta}_{t_j}^{\dag}\hat{\beta}_{t_k}\ket{\Psi}
= \bra{\Psi}\hat{a}_{t_j}^{\dag}\hat{a}_{t_k}\ket{\Psi} + \delta_{j,k}.
\end{eqnarray}
The photon loss process is usually described by a beam splitter model. When one photon is lost with a probability $p$, the mode $\hat{a}_{f}$ is mixed with a vacuum mode $\hat{a}_{f}^{(v)}$ by a beam splitter whose transmittance is $1-p\ (0<p<1)$,
\begin{eqnarray}
\hat{A}_{f} = \sqrt{1-p}\hat{a}_{f}+\sqrt{p}\hat{a}_{f}^{(v)} .
\end{eqnarray}
Similarly, $\hat{\beta}_f$ in Eq. (\ref{eq:beta}) is changed into $\hat{B}_f$ given by
\begin{eqnarray}
\hat{B}_{f} = \sqrt{1-p}\hat{a}_{f}+\sqrt{p}\hat{a}_{f}^{(v1)} -\hat{a}_{f^*}^{\dag (v2)} ,
\end{eqnarray}
where $\hat{a}_{f}^{(v1)}$ and $\hat{a}_{f^*}^{\dag (v2)}$ are vacuum terms due to the photon loss and dual-homodyne measurement. Photon loss degrades pure states $\ket{\Psi}$ into mixed states $\hat{\rho}$, then the $(j,k)$ component of $C_t$ in Eq. (\ref{eq:ct}) is given by
\begin{eqnarray}
C_{t,jk}\left(\hat{\rho}\right)
&=& {\rm tr}\left[ \hat{\rho}\ \hat{B}_{t_j}^{\dag}\hat{B}_{t_k} \right] \nonumber \\
&=&\bra{0_{v_1}}\bra{0_{v_2}}\bra{\Psi}\hat{B}_{t_j}^{\dag}\hat{B}_{t_k}\ket{\Psi}\ket{0_{v_1}}\ket{0_{v_2}} \nonumber  \\
&=& (1-p)\bra{\Psi}\hat{a}_{t_j}^{\dag}\hat{a}_{t_k}\ket{\Psi} + \delta_{j,k} \ .
\end{eqnarray}
Thus, we get
\begin{eqnarray}
C_{t}(\hat{\rho}) = (1-p)C_{t}\left(\ket{\Phi}\bra{\Phi}\right) + pI,
\end{eqnarray}
where $I$ is an $M$-dimensional identity matrix. When $C_t\left(\ket{\Psi}\bra{\Psi}\right)$ is diagonalized by $E$ like Eq. (\ref{eq:diagonalizationE}), $C_{t}(\hat{\rho})$ is also diagonalized by $E$ as follows,
\begin{eqnarray}
E C_{t}(\hat{\rho}) E^{\dag} &=& {\rm diag} \left[ (1-p)\bar{n}_{e_1}+1, \cdots, (1-p)\bar{n}_{e_M}+1 \right] \nonumber \\
&\equiv& {\rm diag} \left[ \bar{N}_{e_1}+1, \cdots, \bar{N}_{e_M}+1 \right]
\end{eqnarray}
Therefore, photon loss only changes $\{ \bar{n}_{e_j} \}_{j=1}^M$ into $\{ \bar{N}_{e_j} \}_{j=1}^M$, and we can assume CPCA gives the same eigenfunctions $\{ e_j(t) \}_{j=1}^M$ in pure state case and mixed state case.

Let us discuss how photon loss affects our analysis method. In single-temporal-mode state analysis, what we want is the eigenfunction $e_1(t)$ as explained in Sec. \ref{2c}, thus photon loss doesn't affect the analysis procedure.

In Sec. \ref{2d}, we calculated $f_1(t), f_2(t)$ from $e_1(t),e_2(t)$ and matrix $D$ given by Eqs. (\ref{eq:linsp2}) and (\ref{Adeg}). We can still obtain $e_1(t),e_2(t)$, but we need slight modification about $D$. We introduced a $4$-th order moment $q\mathrm{e}^{i\theta}$ in Eq. (\ref{eq:qei}). When photon loss exists, what we actually obtain is given by
\begin{eqnarray}
q'\mathrm{e}^{i\Theta}
\equiv {\rm tr}\left[\hat{\rho}\hat{B}_{e_1}^{\dag\ 2}\hat{B}_{e_2}^2\right]
= (1-p)^2q\mathrm{e}^{i\theta} .
\end{eqnarray}
Thus, the phase of the moment is not affected. Because $\bar{N}_{e_1}+\bar{N}_{e_2}=2(1-p)$, we can modify the norm of the moment,
\begin{eqnarray}
Q \equiv \frac{4q'}{\left( \bar{N}_{e_1}+\bar{N}_{e_2} \right)^2}. \label{eq:newq}
\end{eqnarray}
Then, Eqs. (\ref{eq:linsp2}) and (\ref{Adeg}) are modified as follows,
\begin{eqnarray}
D =
\frac{1}{\sqrt{\bar{N}_{e_1}^{\frac{1}{2}}+\bar{N}_{e_2}^{\frac{1}{2}}}}\left(
    \begin{array}{cc}
      \bar{N}_{e_1}^{\frac{1}{4}} \  & i \ \bar{N}_{e_2}^{\frac{1}{4}}\ \ \mathrm{e}^{\frac{i}{2}\Theta} \\
      \bar{N}_{e_1}^{\frac{1}{4}} \  & -i \ \bar{N}_{e_2}^{\frac{1}{4}} \ \mathrm{e}^{\frac{i}{2}\Theta}
    \end{array}
  \right) , \label{eq:finalD}
\end{eqnarray}
\begin{eqnarray}
D =
\left(
    \begin{array}{cc}
      \sqrt{1+\sqrt{1-Q}} \  \ & \mp i \sqrt{1-\sqrt{1-Q}}\ \mathrm{e}^{i\frac{\Theta}{2}} \\
      \sqrt{1-\sqrt{1-Q}} \  \ & \pm i \sqrt{1+\sqrt{1-Q}}\ \mathrm{e}^{i\frac{\Theta}{2}}
    \end{array}
  \right) . \label{eq:lossdeg}
\end{eqnarray}
We can decide the sign of the Eq. (\ref{eq:lossdeg}) by the phase of another moment ${\rm tr}\left[\hat{\rho}\hat{B}_{e_1}^{\dag\ 2}\hat{B}_{e_1}\hat{B}_{e_2}\right]$, which is also not affected by photon loss. Therefore, we can calculate the functions $f_1(t),f_2(t)$ experimentally even when photon loss exists.

In this section, we showed photon loss doesn't affect our TMF estimation essentially. In the next section, we demonstrate these methods.


\section{Experiment}
\subsection{Heralded creation of optical non-Gaussian states\label{3a}}
High purity non-Gaussian states have been created by heralded scheme \cite{Lvovsky.prl(2001),Lvovsky.prl(2002),Yukawa.optexp(2013),Dakna.pra(1997),Ourjoumtsev.science(2006)}. In this method, entangled two modes (idler and signal modes) are prepared and photon detection in the idler mode heralds non-Gaussian states in the signal mode. So far, such states as single-photon states, superposition of photon number states, and Schr\"{o}dinger's cat states have been created \cite{Lvovsky.prl(2001),Lvovsky.prl(2002),Yukawa.optexp(2013),Dakna.pra(1997),Ourjoumtsev.science(2006)}. These created states are defined in wave packet temporal modes, called time-bin modes. The envelopes of the wave packets give the TMF of the modes.

In this scheme, we can engineer the TMFs by the configuration of idler path. Especially, TMFs can be complex when we put an asymmetric Mach-Zehnder interferometer in idler path \cite{Takeda.nature(2013)}. Therefore, heralded creation of non-Gaussian states using interferometers is a good way to demonstrate CPCA. To verify our temporal mode estimation method, we conduct 3 types of heralding experiments with an interferometer. First one is analysis of time-bin qubit as the simplest example. Second one is analysis of what we call {\it dual-rail cat qubit}, qubit consisting of Schr\"{o}dinger's cat states. This experiment shows our method's ability to deal with phase-sensitive and multi-photon states in a single temporal mode. Last one is analysis of time-bin qutrit containing two photons to verify our dual-temporal-mode analysis explained in Sec. \ref{2d}. In the following, we explain how these qubits and qutrits are related to our complex TMF estimation method.

Generation of time-bin qubits and qutrits has already been realized in Refs. \cite{Takeda.nature(2013),Yoshikawa.qph}.
In those studies, the idler and signal modes are realized by two-mode squeezed vacuum emitted from non-degenerate optical parametric oscillator (OPO). One- (two-)photon detection after interferometer(s) in the idler mode heralds time-bin qubits (qutrits) in the signal mode. Time-bin qubits \cite{Takeda.nature(2013)} are generally recognized as two-temporal-mode states, where one photon is distributed in two orthogonal time-bin modes $w_1$ and $w_2$. They are described as $p_1\ket{1_{w_1}}+p_2\ket{1_{w_2}}\equiv p_1\ket{1_{w_1},0_{w_2}}+p_2\ket{0_{w_1},1_{w_2}}$. Here, TMFs $w_1(t)$ and $w_2(t)$ are real functions like in Ref. \cite{Morin.prl(2013)}. By using Eq. (\ref{eq:conb}), these states are transformed as
\begin{eqnarray}
p_1\ket{1_{w_1},0_{w_2}}+p_2\ket{0_{w_1},1_{w_2}}
&=& \left( p_1\hat{a}_{w_1}^{\dag}+ p_2\hat{a}_{w_2}^{\dag} \right) \ket{\tilde{0}} \nonumber \\
&=& \hat{a}_{p_1w_1+p_2w_2}^{\dag}\ket{\tilde{0}} \ . \label{eq:obqubit}
\end{eqnarray}
Thus time-bin qubits are single-temporal-mode single-photon states, whose TMF $p_1w_1(t)+p_2w_2(t)$ are complex function because $p_1,p_2 \in \mathbb{C}$. CPCA can estimate this kind of complex TMFs.

Similarly, time-bin qutrits \cite{Yoshikawa.qph}, two photons distributed in $w_1$ and $w_2$, are described as
\begin{eqnarray}
&&q_1\ket{2_{w_1},0_{w_2}}+q_2\ket{1_{w_1},1_{w_2}}+q_3\ket{0_{w_1},2_{w_2}} \nonumber \\
&=& \frac{1}{\sqrt{1+\abs{r_1^*r_3+r_2^*r_4}^2}} \hat{a}_{r_1w_1+r_2w_2}^{\dag}\hat{a}_{r_3w_1+r_4w_2}^{\dag}\ket{\tilde{0}}, \label{eq:obqutrit}
\end{eqnarray}
as explained in Sec. \ref{2d}. Equation (\ref{eq:obqutrit}) has the same form as Eq. (\ref{eq:trit}), thus we can estimate $f_1(t) = r_1w_1(t)+r_2w_2(t)$ and $f_2(t) = r_3w_1(t)+r_4w_2(t)$, both are the complex functions. We can decide the coefficients $p_1,p_2$ and $r_1,r_2,r_3,r_4$ arbitrarily by changing the power ratio and relative phase of the beams in two arms of the interferometer.

On top of that, we generate dual-rail cat qubits by heralded scheme. Here, the signal mode is one-mode squeezed vacuum emitted from a degenerate OPO. The idler mode is weakly tapped from the signal mode by a beam splitter. We use the same interferometer as time-bin qubit experiments in the idler path. One photon detection in idler mode is recognized as a photon subtraction from the squeezed vacuum, thus the state heralded in the signal mode is given by
\begin{eqnarray}
\left( s_1\hat{a}_{w_1}+s_2\hat{a}_{w_2} \right)\hat{S}_r(w_1)\hat{S}_r(w_2)\ket{\tilde{0}}, \label{eq:subtraction}
\end{eqnarray}
where a squeezing operator of a temporal mode $f$ is given by
\begin{eqnarray}
\hat{S}_r(f) = \exp{\frac{r}{2}\left( \hat{a}_f^{\dag \ 2}-\hat{a}_f^2 \right)}.
\end{eqnarray}
Note that in Eq. (\ref{eq:subtraction}), squeezing operation in the temporal modes orthogonal to $w_1$ and $w_2$ is ignored for simplicity. By choosing proper $r$, photon-subtracted squeezed state becomes very similar to a Schr\"{o}dinger's cat state \cite{Dakna.pra(1997)}. Thus, Eq. (\ref{eq:subtraction}) is a qubit described by
\begin{eqnarray}
s_1\ket{{\rm Cat}_{w_1},{\rm Squeeze}_{w_2}}+s_2\ket{{\rm Squeeze}_{w_1},{\rm Cat}_{w_2}}.
\end{eqnarray}
Note that the basis states of the qubits are orthogonal because cat states (squeezed states) have only odd (even) photon number components. We can decide $s_1,s_2 \in \mathbb{C}$ by the interferometer arbitrarily in the same way as time-bin qubits case. For example, when $s_1=s_2=1/\sqrt{2}$, Eq. (\ref{eq:subtraction}) is given by
\begin{eqnarray}
\hat{a}_{\frac{w_1+w_2}{\sqrt{2}}}\hat{S}_r\left(\frac{w_1+w_2}{\sqrt{2}}\right)\hat{S}_r\left(\frac{w_1-w_2}{\sqrt{2}}\right)\ket{\tilde{0}}. \label{eq:cat}
\end{eqnarray}
Thus, it is a single-temporal-mode Schr\"{o}dinger's cat state in $\left( w_1+w_2 \right)/\sqrt{2}$. Similar mode transformation of squeezed operation is seen in \cite{Takahashi(2008)}. On the other hand, when we consider complex temporal modes, we need introduce a two-mode squeezing operator. When $s_1=1/\sqrt{2},s_2=-i/\sqrt{2}$, Eq. (\ref{eq:subtraction}) is
\begin{eqnarray}
\hat{a}_{\frac{w_1-iw_2}{\sqrt{2}}}\hat{S}_r^{(2)}\left(\frac{w_1+iw_2}{\sqrt{2}},\frac{w_1-iw_2}{\sqrt{2}}\right)\ket{\tilde{0}}, \label{eq:psepr}
\end{eqnarray}
where a two-mode squeezing operator of orthogonal modes $f_1,f_2$ is given by
\begin{eqnarray}
\hat{S}_r^{(2)}(f_1,f_2) = \exp{r\left( \hat{a}_{f_1}^{\dag}\hat{a}_{f_2}^{\dag}-\hat{a}_{f_1}\hat{a}_{f_2} \right)}.
\end{eqnarray}
In photon number basis, Eq. (\ref{eq:psepr}) is given by \cite{Walls(1994)}
\begin{eqnarray}
&&\hat{a}_{\frac{w_1-iw_2}{\sqrt{2}}} \frac{1}{\cosh{r}} \sum_{n=0}^{\infty}(\tanh{r})^n \ket{n_{\frac{w_1+iw_2}{\sqrt{2}}},n_{\frac{w_1-iw_2}{\sqrt{2}}}} \nonumber \\
&=& \frac{1}{\cosh{r}} \sum_{n=0}^{\infty}\sqrt{n+1}\ (\tanh{r})^{n+1} \ket{n+1_{\frac{w_1+iw_2}{\sqrt{2}}},n_{\frac{w_1-iw_2}{\sqrt{2}}}}. \label{eq:psepr2}
\end{eqnarray}
When $r \to \infty$, the created state is a photon-subtracted Einstein-Podolsky-Rosen (EPR) state. Subtraction makes the average photon number larger in $\left( w_1-iw_2 \right)/\sqrt{2}$ and especially in $\left( w_1+iw_2 \right)/\sqrt{2}$. This is a kind of entanglement purification similar to Ref. \cite{Takahashi(2010)}.

The modes $\left( w_1+iw_2 \right)/\sqrt{2}$ and $\left( w_1-iw_2 \right)/\sqrt{2}$ are orthogonal, and satisfy
\begin{eqnarray}
\braket{\hat{\beta}_{\frac{w_1+iw_2}{\sqrt{2}}}^{\dag}\hat{\beta}_{\frac{w_1-iw_2}{\sqrt{2}}}} = \braket{\hat{\beta}_{\frac{w_1-iw_2}{\sqrt{2}}}^{\dag}\hat{\beta}_{\frac{w_1+iw_2}{\sqrt{2}}}} = 0.
\end{eqnarray}
Therefore, it is expected that CPCA gives $e_1(t)=\left( w_1(t)+iw_2(t) \right)/\sqrt{2}$ and $e_2(t)=\left( w_1(t)-iw_2(t) \right)/\sqrt{2}$.

\subsection{Experimental setup\label{3b}}
\begin{figure}[tb]
\centering
\includegraphics[width=\linewidth]{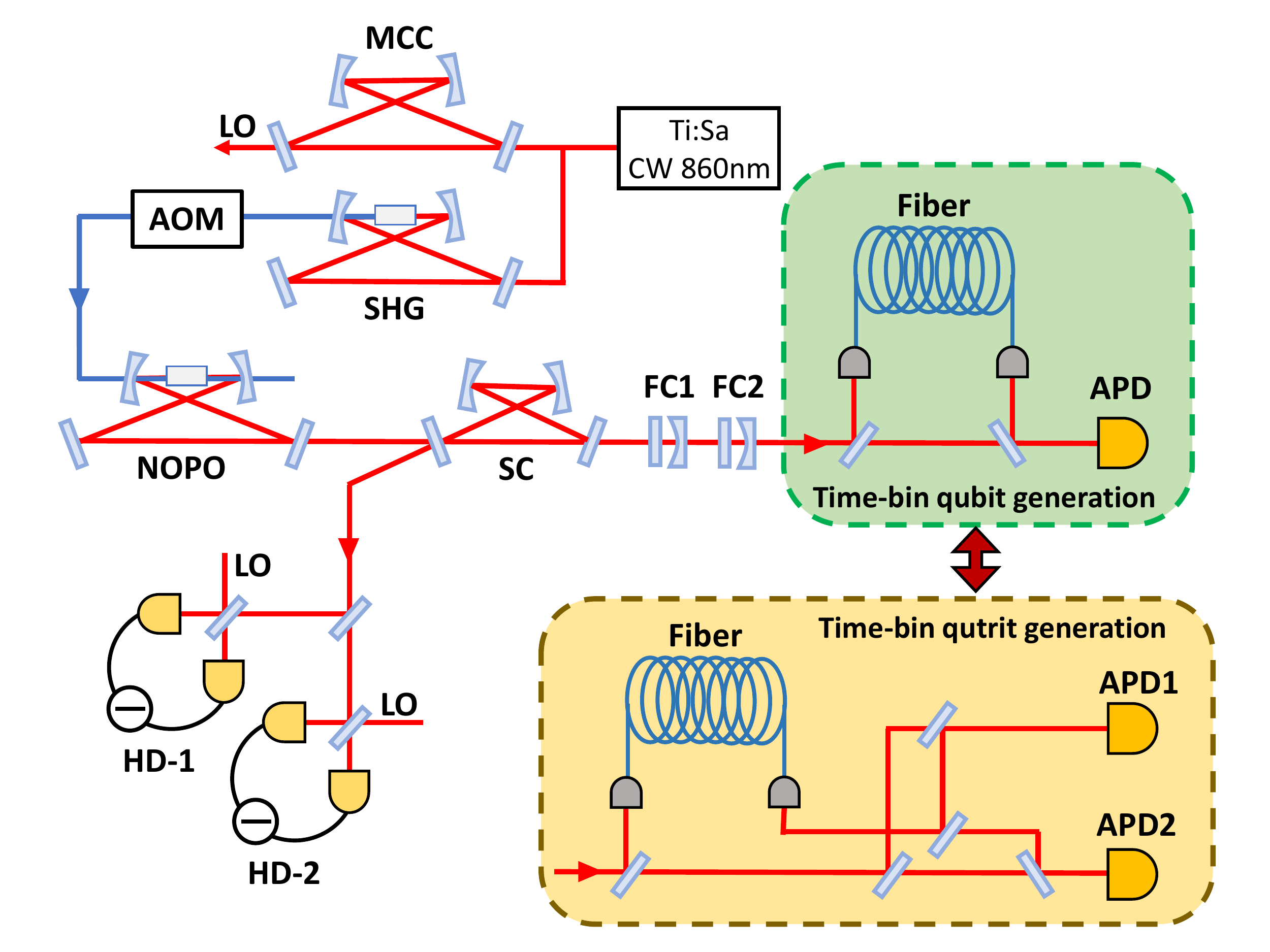}
\caption{
Experimental setup for the heralded creation of time-bin qubits $p_1\ket{1_{w_1},0_{w_2}}+p_2\ket{0_{w_1},1_{w_2}}$, and time-bin qutrits $q_1\ket{2_{w_1},0_{w_2}}+q_2\ket{1_{w_1},1_{w_2}}+q_2\ket{0_{w_1},2_{w_2}}$. Ti:Sa denotes titanium sapphire laser, CW continuous wave, SHG second harmonic generator, MCC mode cleaning cavity, AOM acousto-optical modulator, NOPO non-degenerate optical prametric oscillator, SC splitting cavity, FC filter cavity, APD avalanche photo diode, LO local oscillator, and HD homodyne detector. In the photon subtraction experiments, AOM is removed and SC is replaced by $97\%$ reflection beam splitter in the time-bin qubit generation setup.
}
\label{fig:setup_qubit}
\end{figure}

The schematic diagram of time-bin qubit and qutrit generation is shown in Fig. \ref{fig:setup_qubit}. This setup is the same as Refs. \cite{Takeda.nature(2013),Yoshikawa.qph}. The light source of the experiment is a CW Ti:Sapphire laser whose wavelength is $860\ {\rm nm}$. A bow-tie shaped non-degenerate OPO is used to generate two-mode squeezed vacuum states. The cavity of the OPO has $16$ MHz of full width at half maximum (FWHM) and $600$ MHz of free spectrum range (FSR). Inside the cavity, $10$ mm length periodically poled ${\rm KTiOPO_4}$ crystal is placed as a nonlinear optical medium. The pump beam of the OPO is produced from a second harmonic generator (SHG), which consists of a bow-tie shaped cavity and $10$ mm length ${\rm KNbO_3}$ crystal. The pump beam is given $600$ MHz (one FSR) frequency shift by an acousto-optical modulator (AOM).

Signal and idler modes of the two-mode squeezed vacuum states have different frequency and are divided into two optical paths by a splitting cavity. The idler mode passes through two Fabry-P\'{e}rot filter cavities to filter out unwanted non-degenerate modes. When the OPO is weakly pumped, photon detection by a silicon avalanche photo diode (APD) in the idler mode heralds single-photon states in the signal mode. In order to create time-bin qubits, we construct an asymmetric Mach-Zehnder interferometer between the filter cavities and the APD. The idler field in the longer arm of the interferometer is given time delay against the idler field in the shorter arm, thus time-shifted idler fields interfere before photon detection. This interference enables the photon detection of APD to herald time-bin superposition states. The longer arm of interferometer is implemented by an about $50$ m optical fiber, which is long enough to regard the heralded two time-bins as orthogonal. Arbitrary time-bin qubits can be created by changing the power ratio and relative phase of two beams in the interferometer. In the case of creation of time-bin qutrits, we combine two asymmetric Mach-Zehnder interferometers and two APDs. In this case, simultaneous photon detection at two APDs heralds time-bin qutrits in the signal mode. 

In dual-rail cat qubit generation, we remove the AOM to use the OPO as a degenerate OPO to generate one-mode squeezed vacuum. The splitting cavity is replaced by a $97\%$ reflection beam splitter. The interferometer in the idler path is same as time-bin qubit setup. We generate single-temporal-mode Schr\"{o}dinger's cat states and photon-subtracted EPR states by changing the relative phase of the beams in the two arms of the interferometer.

All these created states are detected by CW dual homodyne measurement. The transversal mode of the LO beams for the homodyne measurement is set to ${\rm TEM_{00}}$ by a bow-tie shaped mode cleaning cavity. The sampling rate of data acquisition is $1$GHz and one data frame contains $1,500$ points ($T=1.5 {\rm \mu s} \ , \  M = 1500$). Each state is measured $20,000$ times to construct the matrix $C_t$ introduced in Eq. (\ref{eq:ct}). In this case, $C_t$ is a $1,500\times 1,500$ dimensional matrix. Measured values are filtered by 2nd order LC high-pass-filter and digital low-pass-filter to filter out the effect of large signal noise at DC and gain peaking of homodyne detectors at high frequency. The cut-off frequencies are 100 kHz and 14.3 MHz respectively.

Theoretically, time-bin TMF has double decaying exponential profile $\sqrt{\gamma}\mathrm{e}^{-\gamma|t|}$ where $\gamma= 1.1\times 10^8\ $/s as a Fourier counterpart of OPO's Lorentzian frequency spectrum \cite{Anne.pra(2007)}. The low-pass filter effect of two filter cavities and digital filter make the actual TMF $w(t)$ to be round-shaped \cite{Anne.pra(2006)}. On top of that, 50 m optical fiber makes time-shifted superposition of two time-bin modes $w_1(t)$ and $w_2(t)$, where $w_2(t)=w_1(t-\Delta t), \Delta t \approx 250$ ns. This time delay is enough to assume that these two modes are orthogonal, considering the exponential decay of the function $w(t)$ given by $\gamma$. We estimate the mode functions seen in Eqs. (\ref{eq:obqubit}) and (\ref{eq:obqutrit}), thus experimental results are expected to be superposition of $w_1(t)$ and $w_2(t)$, as you can see in the next section.

\subsection{Results\label{3c}}
\subsubsection{Time-bin qubit and dual-rail cat qubit}
\begin{figure*}[t]
\centering
\includegraphics[width=0.9\linewidth]{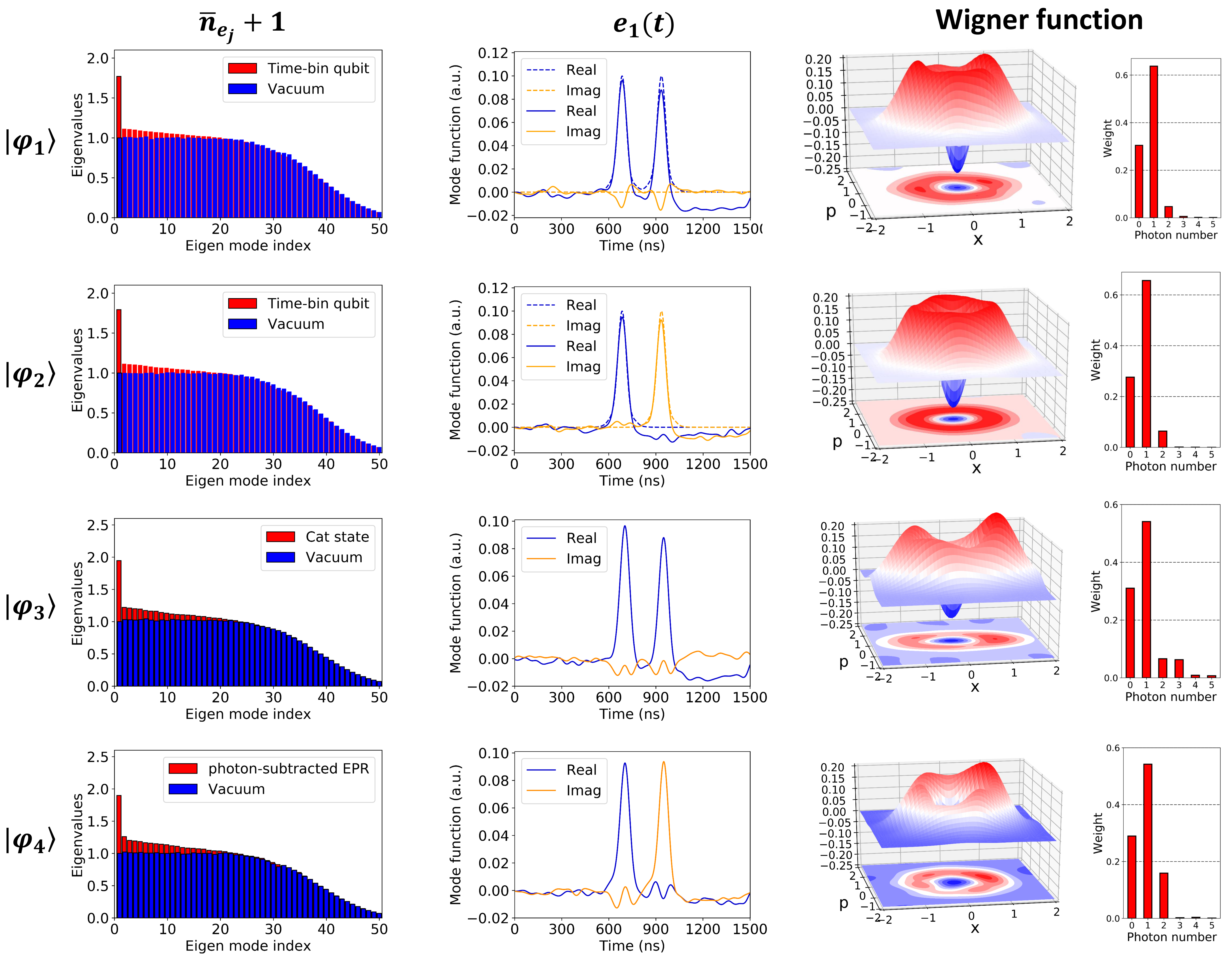}
\caption{
Analysis results of $\ket{\phi_1}$ to $\ket{\phi_4}$. Left: First 50 eigenvalues of matrix $C_t$ are shown in red bar (back), and vacuum state in blue (front). Middle: The first eigenfunctions $e_1(t)$ are shown in real lines. Blue and orange lines show the real and imaginary part of $e_1(t)$ respectively. As for $\ket{\phi_1}$ and $\ket{\phi_2}$, theoretical predictions are shown in broken lines. Right: Wigner functions and photon number distributions of $e_1$.
}
\label{fig:photoncat}
\end{figure*}
We show the analysis results of 2 types of time-bin qubits given by
\begin{eqnarray}
\ket{\phi_1} &=& \frac{1}{\sqrt{2}}\left( \ket{1_{w_1},0_{w_2}}+\ket{0_{w_1},1_{w_2}} \right) =  \hat{a}_{\frac{w_1+w_2}{\sqrt{2}}}^{\dag}\ket{\tilde{0}}, \\
\ket{\phi_2} &=& \frac{1}{\sqrt{2}}\left( \ket{1_{w_1},0_{w_2}}+i\ket{0_{w_1},1_{w_2}} \right) =  \hat{a}_{\frac{w_1+iw_2}{\sqrt{2}}}^{\dag}\ket{\tilde{0}}.
\end{eqnarray}
and 2 types of dual-rail cat qubit given by
\begin{eqnarray}
\ket{\phi_3} &=& \frac{1}{\sqrt{2}}\ket{{\rm Cat}_{w_1},{\rm Squeeze}_{w_2}}+\frac{1}{\sqrt{2}}\ket{{\rm Squeeze}_{w_1},{\rm Cat}_{w_2}} \nonumber \\
&\propto& \hat{a}_{\frac{w_1+w_2}{\sqrt{2}}}\hat{S}_r\left(\frac{w_1+w_2}{\sqrt{2}}\right)\hat{S}_r\left(\frac{w_1-w_2}{\sqrt{2}}\right)\ket{\tilde{0}}, \\
\ket{\phi_4} &=& \frac{1}{\sqrt{2}}\ket{{\rm Cat}_{w_1},{\rm Squeeze}_{w_2}}-\frac{i}{\sqrt{2}}\ket{{\rm Squeeze}_{w_1},{\rm Cat}_{w_2}} \nonumber \\
&\propto& \hat{a}_{\frac{w_1-iw_2}{\sqrt{2}}}\hat{S}_r^{(2)}\left(\frac{w_1+iw_2}{\sqrt{2}},\frac{w_1-iw_2}{\sqrt{2}}\right)\ket{\tilde{0}}.
\end{eqnarray}
Figure \ref{fig:photoncat} shows the CPCA results of those 4 states. The left row shows first $50$ eigenvalues of $C_t$, that is, $\{ \bar{N}_{e_j}+1 \}_{j=1}^{50}$. You can see that the first eigenvalue is outstanding in each case. Other eigenvalues, which correspond to thermal states about $\ket{\phi_1},\ket{\phi_2}$ and squeezed states about $\ket{\phi_3},\ket{\phi_4}$, are slightly larger than vacuum states. Due to the digital low-pass-filter, those modes containing high frequency components have small eigenvalues. It follows that the eigenvalues go below $1$ as the mode index increases.
\begin{table}[t]
  \begin{tabular}{|c|c|c|} \hline
    state & expected TMF & mode match \\ \hline \hline
    $\ket{\phi_1}$ & $e_1(t)\propto w_1(t)+w_2(t)$ & 0.863 \\ \hline
    $\ket{\phi_2}$ & $e_1(t)\propto w_1(t)+iw_2(t)$ & 0.912 \\ \hline
    $\ket{\phi_3}$ & $e_1(t)\propto w_1(t)+w_2(t)$ & 0.862 \\ \hline
    $\ket{\phi_4}$ & $e_1(t)\propto w_1(t)+iw_2(t)$ & 0.913 \\
     & $e_2(t)\propto w_1(t)-iw_2(t)$ & 0.630 \\ \hline
    $\ket{\phi_5}$ & $f_1(t)\propto w_1(t)+iw_2(t)$ & 0.956 \\
     & $f_2(t)\propto w_1(t)-iw_2(t)$ & 0.946 \\ \hline
    $\ket{\phi_6}$ & $f_1(t)\propto w_1(t)+\mathrm{e}^{i\frac{\pi}{4}}w_2(t)$ & 0.827 \\
     & $f_2(t)\propto w_1(t)+\mathrm{e}^{-i\frac{\pi}{4}}w_2(t)$ & 0.870 \\ \hline
  \end{tabular}
\caption{Left row: analyzed states. Middle row: theoretically expected TMFs. Right row: mode match between estimated TMFs and theoretical predictions. }
\label{tab:1}
\end{table}

The middle row of Fig. \ref{fig:photoncat} shows $e_1(t)$ plotted in real lines. Blue and orange lines correspond to real and imaginary part of $e_1(t)$. We can see that $e_1(t)$ consists of two time-bins ($w_1(t)$ and $w_2(t)$). The first time-bins appear in real part, and the second time-bins appear in real or imaginary part of $e_1(t)$. These represent the relative phases of the superposition of two time-bins. As for $\ket{\phi_1}$ and $\ket{\phi_2}$, we show the theoretical predictions in broken lines. The experimental results capture the features of theoretical predictions well. Table \ref{tab:1} shows theoretically expected TMFs and mode match of the experimental results and the theoretical predictions. The mode matches are reasonably high, but some mismatch comes from mainly two reasons. One is imperfection of the interferometer in the idler path. The mode match goes down when the power ratio or phase of the beams in the two arms of the interferometer are not set correctly. The other reason is the effect of high-pass-filter after the homodyne detectors. Especially in $\ket{\phi_1}$ and $\ket{\phi_3}$, you can see that the long tail of time-bin due to the filter makes the mode match worse.
\begin{figure*}[t]
\centering
\includegraphics[width=0.9\linewidth]{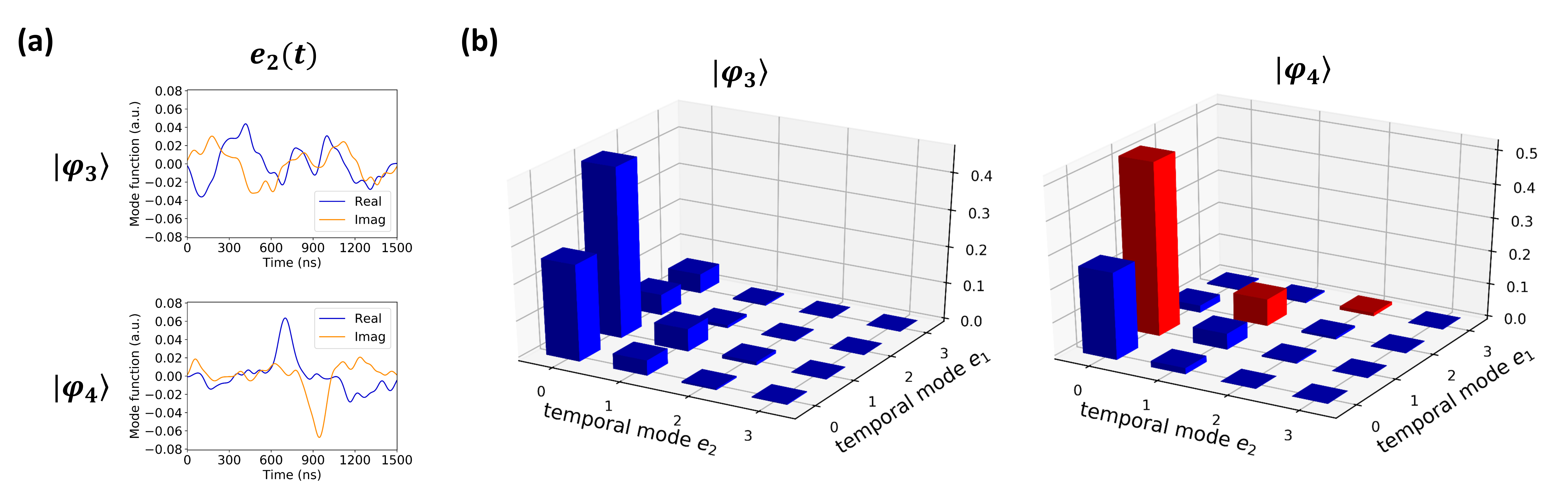}
\caption{
(a)Second eigenfunction $e_2(t)$ of $\ket{\phi_3}$ and $\ket{\phi_4}$.\ (b)Two-mode photon number distribution about $e_1$ and $e_2$. Red bars show expected photon number correlation of $\ket{\phi_4}$.
}
\label{fig:r7}
\end{figure*}
\begin{figure*}[t]
\centering
\includegraphics[width=0.9\linewidth]{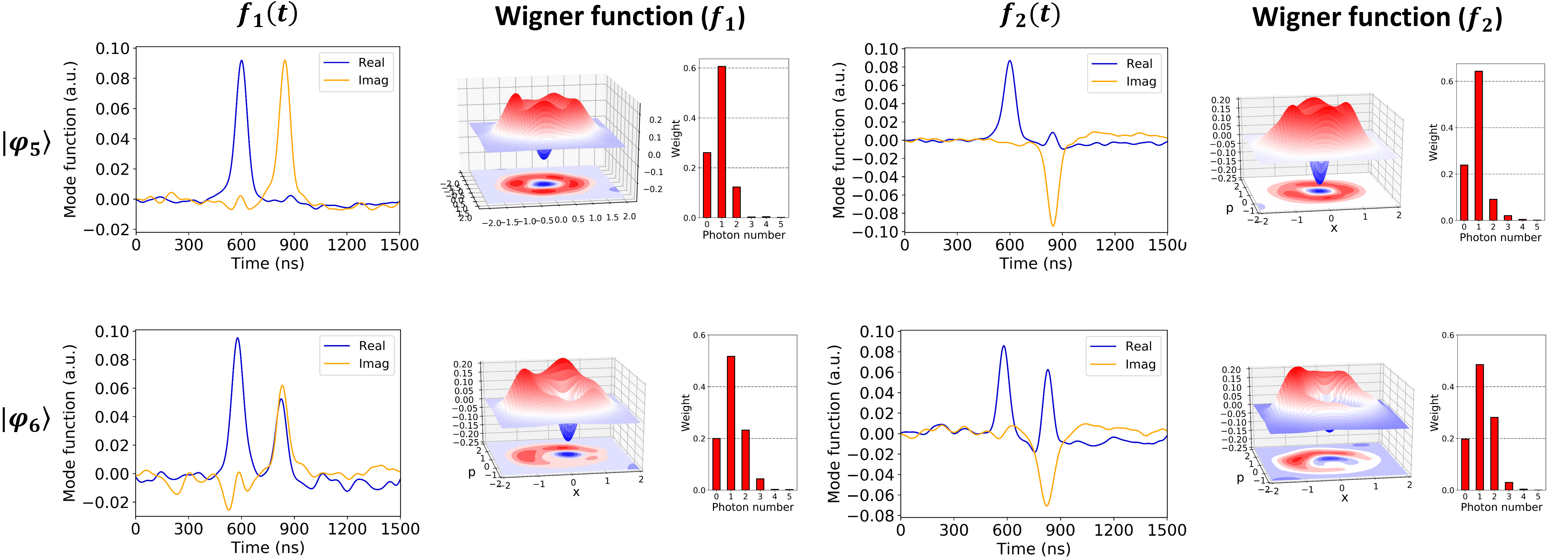}
\caption{
Estimated $f_1(t),f_2(t)$ of time-bin qutrits $\ket{\phi_5}$ and $\ket{\phi_6}$. Wigner function and photon number distribution of each modes are also shown.
}
\label{fig:tritnew}
\end{figure*}

The right row of Fig. \ref{fig:photoncat} shows Wigner functions and photon number distributions of temporal mode $e_1$. Wigner functions have negative values, thus these states are non-Gaussian states having high non-classicality. $\ket{\phi_1}$ and $\ket{\phi_2}$ have more than 60 percent single-photon components and their Wigner functions have rotational symmetry. Thus, they are high purity single-photon states as expected. The Wigner function of $\ket{\phi_3}$ are squeezed in $p$ direction and not rotationally symmetric, which is one prominent feature of Schr\"{o}dinger's cat state. From Eq. (\ref{eq:psepr2}), $\ket{\phi_4}$ is a mixed state of $\ket{1_{e_1}},\ket{2_{e_1}},\cdots$ in $e_1$, thus their Wigner function is expected to be rotationally symmetric. Actually their Wigner function is rotationally symmetric. These results confirm that the experiments have been carried out successfully.

Figure \ref{fig:r7}(a) shows $e_2(t)$ of $\ket{\phi_3}$ and $\ket{\phi_4}$. $e_2(t)$ of $\ket{\phi_3}$ has fluctuating wave form. It is chosen to have the largest average photon number in background squeezed states. On the other hand, $e_2(t)$ of $\ket{\phi_4}$ consists of two time-bins as explained in Eq. (\ref{eq:psepr2}). In this case, however, the wave form is a little vague and mode match is not so high compared to other cases as shown in Table \ref{tab:1}. This may be because $\bar{N}_{e_2}$ is only slightly larger than $\bar{N}_{e_3}$, thus it becomes difficult to separate the expected mode from background. As Eq. (\ref{eq:psepr2}) tells, the temporal modes $e_1$ and $e_2$ of $\ket{\phi_4}$ are entangled. You can see the photon number correlation given by $\ket{n+1_{e_1}}\ket{n_{e_2}}$ in two-mode photon number distribution about $e_1$ and $e_2$ in Fig. \ref{fig:r7}(b). The correlation coefficient of $n_{e_1}$ and $n_{e_2}$ is $r=0.498$. On the other hand, $\ket{\phi_3}$ has little photon number correlation between $e_1$ and $e_2$ ($r=-0.015$) as shown in Fig. \ref{fig:r7}(b).

These results show that CPCA can access the complex TMFs of various single-temporal-mode non-Gaussian states.


\subsubsection{Results of time-bin qutrit experiments}
We analyze $2$ types of time-bin qutrits given by
\begin{eqnarray}
\ket{\phi_5} &=& \frac{1}{\sqrt{2}}\left( \ket{2_{w_1},0_{w_2}}+\ket{0_{w_1},2_{w_2}} \right) =  \hat{a}_{\frac{w_1+iw_2}{\sqrt{2}}}^{\dag}\hat{a}_{\frac{w_1-iw_2}{\sqrt{2}}}^{\dag}\ket{\tilde{0}}, \\
\ket{\phi_6} &=& \frac{1}{\sqrt{3}}\left( \ket{2_{w_1},0_{w_2}}+\ket{1_{w_1},1_{w_2}}+\ket{0_{w_1},2_{w_2}} \right) \nonumber \\ 
&=&  \hat{a}_{\frac{w_1+\mathrm{e}^{i\frac{\pi}{4}}w_2}{\sqrt{2}}}^{\dag}\hat{a}_{\frac{w_1+\mathrm{e}^{-i\frac{\pi}{4}}w_2}{\sqrt{2}}}^{\dag}\ket{\tilde{0}}.
\end{eqnarray}
We can calculate $f_1(t)$ and $f_2(t)$ of these dual-temporal-mode two-photon states by CPCA results in the way explained in Sec. \ref{2d}. Figure \ref{fig:tritnew} is the analysis result showing calculated $f_1(t),f_2(t)$, and Wigner functions and photon number distributions of those modes. The theoretical TMFs and mode match are written in table \ref{tab:1}. Note that in  ideal case, $\ket{\phi_5}$ satisfies $\braket{f_1,f_2}=0$ and $\bar{N}_{e_1}=\bar{N}_{e_2}$, thus we should use Eq. (\ref{eq:lossdeg}) to calculate $f_1(t)$ and $f_2(t)$. However, imperfection of experimental conditions makes $\bar{N}_{e_1}>\bar{N}_{e_2}$.
Thus, we calculate $f_1(t),f_2(t)$ using Eq. (\ref{eq:finalD}) for both $\ket{\phi_5}$ and $\ket{\phi_6}$.

Photon number distributions in Figure \ref{fig:tritnew} have larger weight in two-photon distribution compared to time-bin qubits in Fig. \ref{fig:photoncat}. One reason is that $f_1(t)$ and $f_2(t)$ are not orthogonal with inner product $\abs{\braket{f_1, f_2}}=0.046$ for $\ket{\phi_5}$ and $\abs{\braket{f_1, f_2}}=0.475$ for $\ket{\phi_6}$. This means that some components in mode $f_1$ is mixed into the mode $f_2$, and vice versa. This leads to larger multi-photon components in each modes. Another reason is that the two-mode squeezed vacuum contains larger multi-photon components than time-bin qubit's case because we use higher pumping condition to have enough count rate of simultaneous photon detection at two APDs. Like above, we can analyze arbitrary time-bin qutrits in the way introduced in Sec. \ref{2d}.

We analyzed experimentally created time-bin qubits, dual-rail cat qubits, and time-bin qutrits. These results show that CPCA method enables us to access complex temporal-mode structures of optical non-Gaussian states.

\section{Conclusion\label{4}}
We introduced CPCA, a method to estimate complex TMFs of optical non-Gaussian states. It is based on principal components analysis of complex variables given by continuous-wave dual-homodyne measurement. CPCA can deal with not only arbitrary single-temporal-mode non-Gaussian states, but also arbitrary dual-temporal mode two-photon states $\hat{a}_{f_1}^{\dag}\hat{a}_{f_2}^{\dag}\ket{\tilde{0}}$, which previous methods cannot deal with. We showed that our scheme works in actual situation by analyzing several experimentally non-Gaussian states. CPCA needs only simple experimental setup, two homodyne detectors and continuous-wave local oscillator beam having one frequency. Analysis procedure is also simple; it's basically just a diagonalization of a matrix. Due to the simplicity and capability to characterize wide range of quantum states, our method is a powerful tool in state creation experiments. Estimated TMFs reflect the imperfection of experiments, thus we can utilize the created states with high purity. This achievement would lead to optimization of quantum communication and quantum computation systems.

\section{Acknowledgements}
This work was partly supported by CREST (JPMJCR15N5) and PRESTO (JPMJPR1764) of JST, JSPS KAKENHI, and UTokyo Foundation. K. T., M. O., and T. S. acknowledge financial support from ALPS.


\end{document}